\documentclass[a4paper,fleqn,usenatbib]{mnras}
\usepackage{newtxtext,newtxmath}
\usepackage[T1]{fontenc}
\usepackage[utf8]{inputenc}
\usepackage{ae,aecompl}
\usepackage{graphicx}	
\usepackage{multirow}
\usepackage{booktabs}
\usepackage{multicol}
\usepackage{gensymb}
\usepackage{subcaption}
\usepackage{array} 
\usepackage{natbib}
\usepackage{amsmath}	
\usepackage{amssymb}	
\usepackage{epstopdf}
\newcolumntype{Y}{>{\centering\arraybackslash}X}
\usepackage{scalerel}
\usepackage{tikz}
\usetikzlibrary{svg.path}
\definecolor{orcidlogocol}{HTML}{A6CE39}
\tikzset{
  orcidlogo/.pic={
    \fill[orcidlogocol] svg{M256,128c0,70.7-57.3,128-128,128C57.3,256,0,198.7,0,128C0,57.3,57.3,0,128,0C198.7,0,256,57.3,256,128z};
    \fill[white] svg{M86.3,186.2H70.9V79.1h15.4v48.4V186.2z}
                 svg{M108.9,79.1h41.6c39.6,0,57,28.3,57,53.6c0,27.5-21.5,53.6-56.8,53.6h-41.8V79.1z M124.3,172.4h24.5c34.9,0,42.9-26.5,42.9-39.7c0-21.5-13.7-39.7-43.7-39.7h-23.7V172.4z}
                 svg{M88.7,56.8c0,5.5-4.5,10.1-10.1,10.1c-5.6,0-10.1-4.6-10.1-10.1c0-5.6,4.5-10.1,10.1-10.1C84.2,46.7,88.7,51.3,88.7,56.8z};
  }
}

\newcommand\orcidicon[1]{\href{https://orcid.org/#1}{\mbox{\scalerel*{
\begin{tikzpicture}[yscale=-1,transform shape]
\pic{orcidlogo};
\end{tikzpicture}
}{|}}}}
\hypersetup{bookmarks,bookmarksopen,bookmarksdepth=2}

\title{Earth-size planet formation in the habitable zone of circumbinary stars}

\author[G. O. Barbosa et al.]{
G. O. Barbosa,$^{1,2}$\thanks{E-mail: gerson.barbosa@inpe.br (GOB)}\orcidicon{0000-0002-1147-2519}\,
O. C. Winter,$^{2}$\thanks{E-mail: othon.winter@unesp.br (OCW)}\orcidicon{0000-0002-4901-3289}\,
A. Amarante,$^{2,3,4}$ \thanks{E-mail: andre.amarante@ifsp.edu.br (AA)}\orcidicon{0000-0002-9448-141X}\,
A. Izidoro$^{2}$ \thanks{E-mail: izidoro.costa@gmail.com (AI)}\orcidicon{0000-0003-1878-0634}\,
\newauthor
\hspace{0.15cm}R. C. Domingos$^{5}$ \thanks{E-mail: rita.domingos@unesp.br (RCD)}\orcidicon{0000-0002-0516-0420}\,
E. E. N. Macau$^{1, 6}$ \thanks{E-mail: elbert.macau@unifesp.br (EENM)}\orcidicon{0000-0002-6337-8081}\
\\
$^{1}$National Institute for Space Research (INPE), Laboratório de Computação Aplicada,
             São José dos Campos, SP 12227-010,  Brazil.\\
$^{2}$São Paulo State University (UNESP), Grupo de Dinâmica Orbital e Planetologia, 
             Guaratinguetá, SP 12516-410, Brazil.\\
$^{3}$State University of Mato Grosso do Sul (UEMS), Cassil\^andia, MS 79540-000, Brazil.\\
$^{4}$Federal Institute of Education, Science and Technology of S\~ao Paulo (IFSP), Cubat\~ao, SP 11533-160, Brazil.\\
$^{5}$São Paulo State University (UNESP),
             São João da Boa Vista, SP 13876750, Brazil.\\
$^{6}$Federal University of São Paulo (UNIFESP), Institute for Science and Technology, 
             São José dos Campos, SP 12247-014, Brazil.
}

\date{Accepted 2020 March 16. Received 2020 March 1; in original form 2019 December 20}

\pubyear{2020}

\begin{document}
\label{firstpage}
\pagerange{\pageref{firstpage}--\pageref{lastpage}}
\maketitle

\begin{abstract}
In this work is investigated the possibility of close-binary star systems having Earth-size planets within their habitable zones. First, we selected all known close-binary systems with confirmed planets (totaling 22 systems) to calculate the boundaries of their respective habitable zones (HZ). However, only eight systems had all the data necessary for the computation of the HZ.  Then, we numerically explored the stability within the habitable zones for each one of the eight systems using test particles. From the results, we selected five systems that have stable regions inside the habitable zones (HZ), namely \textit{Kepler-34, 35, 38, 413} and \textit{453}. For these five cases of systems with stable regions in the HZ, we perform a series of numerical simulations for planet formation considering disks composed of planetary embryos and planetesimals, with two distinct density profiles, in addition to the stars and host planets of each system. We found that in the case of \textit{Kepler-34} and \textit{453} systems no Earth-size planet is formed within the habitable zones. Although planets with Earth-like masses were formed in the \textit{Kepler-453}, but they were outside the HZ. In contrast, for \textit{Kepler-35} and \textit{38} systems, the results showed that potentially habitable planets are formed in all simulations. In the case of the \textit{Kepler-413 system}, in just one simulation a terrestrial planet was formed within the habitable zone.
\end{abstract}

\begin{keywords}
planets and satellites: formation -- (stars:) binaries (including multiple): close
\end{keywords}



\section{Introduction}

Today there are around 4,100 exoplanets confirmed in more than 3,000 stellar systems, most of which were discovered by the \textit{Kepler} probe \citep{borucki2010kepler, howell2014k2}. From these systems with planets, around 600 are multiplanetary, that is, with at least two confirmed planets. In addition to recent discoveries in single star systems, planets have also been discovered in multiple (at least three stars) and binary star systems and about 50\% of Sun-type stars are in binary or multiple systems \citep{raghavan2010survey}. In the binary case, there are known 22 stellars systems with planets in circumbinary orbits, i.e., planets around the barycenter of the binary pair, and none of them is a terrestrial (Table \ref{systems}). The effect of the binary companion on the formation of these planets, however, remains unclear. On the other hand, recent works \citep{quintana2006terrestrial, quintana2010terrestrial, haghighipour2007habitable, quintana2007terrestrial} explored the terrestrial planet formation around close binary (CB) and have shown that it is possible to form planets in this context.

Some works \citep{gorlova2006spitzer,trilling2007debris,furlan2007hd} based on spectral radiation characteristics in the infrared region and the radial velocity method have suggested the existence of binary stellar systems with massive protoplanetary disks. Indirect observations suggest the existence of disk material around one of the stars or the components of the system of stars \citep{d1997structure, stapelfeldt1998edge, jensen2003protoplanetary,osorio2003comprehensive, akeson2014circumstellar}. \cite{jensen1997evidence} showed that protoplanetary disks are common around binary stars with less separation than a few astronomical units. Observations of pairs of stars in the main sequence by the \textit{Spitzer} Space Telescope revealed stable circumferential protoplanetary disks around pairs of stars with spacings between 0.04 and 5.31 au in 14 systems \citep{trilling2007debris}, with two of the systems having planets. Although binary systems have only gaseous planets, this evidence demonstrates the possibility of them forming 
terrestrial planets.

\begin{figure}
\centering
\includegraphics*[width=\hsize]{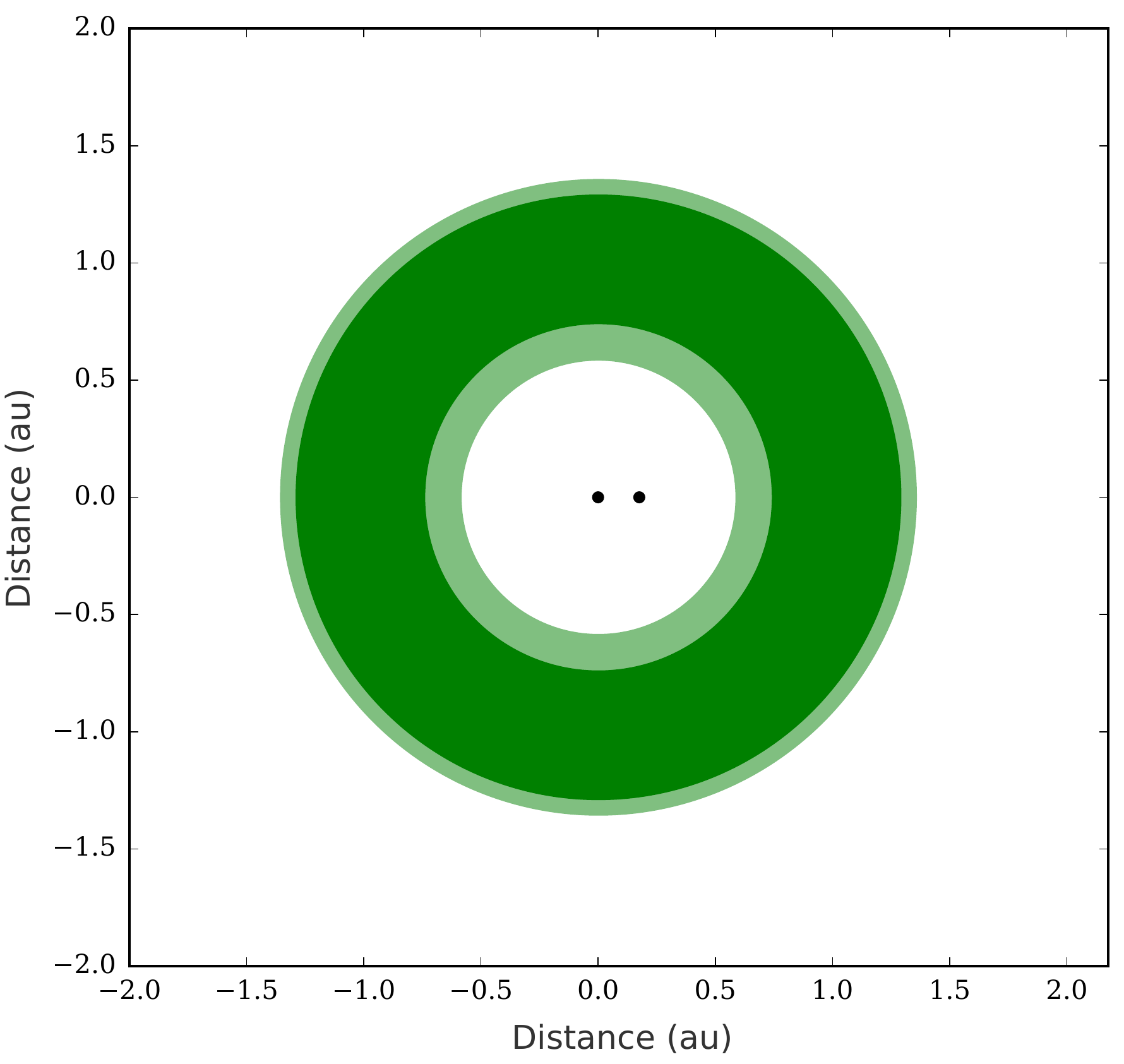}
\caption{Example of a habitable zone of system \textit{Kepler-453} \protect{\citep{haghighipour2013calculating}}. This figure was generated using the interactive website \protect{\url{http://astro.twam.info/hz/}}. The figure is centered on the center of mass of the pair of stars in black. The conservative and extended HZ are in dark green and light green, respectively.}
\label{FigVibStab}
\end{figure}

In \cite{domingos2012mean} it was shown that in regions of the disk close to a binary system, bodies could be captured in mean motion resonances, which according to recent studies \citep{verrier2008hd, farago2010high}, could be a potential mechanism of instability and disk failure. In addition, if the disk is tilted relative to the orbit plane of the binary system, the disk bodies close to the binary system must undergo the hovering effect of the node \citep{verrier2009high, domingos2015planet}, which, depending on the inclination of the disk, should be stronger than the \textit{Kozai-Lidov} effect. Therefore, the node's libration tends to stabilize the system, what results in a stable region. However, the orbits are tilted, what could make the planetary formation process longer when compared to single star systems \citep{domingos2015planet}.

The search for life is one of the great motivators 
of humanity for the exploration of our Solar system and other planets outside it. The knowledge of regions that make possible its existence as we know is fundamental for this search. Habitable zone (HZ) is defined as a region where Earth-like planets could support life. This zone is defined as a circumstellar region where planets with an atmosphere composed of $CO_2-H_2O-N_2$ can sustain liquid water on its surface \citep{huang1959occurrence, hart1978evolution, kasting1993habitable, underwood2003evolution, selsis2007habitable, kaltenegger2011exploring}. In \cite{haghighipour2013calculating}, a generalized model that estimates the limits of HZ for binary systems, shows that they have these well-defined limits, see Figure \ref{FigVibStab}.

In \cite{shevchenko2017habitability} it is shown that CB systems are better able to harbor life, since a system that has two Sun-type stars has a wider HZ than systems of only one star of the same type, moreover, in a biological context, they do not need to have a large Moon-type satellite (as in the case of Earth) to be habitable. This favors the possibility that a Earth-type planet has life given the extremely low probability of it having a Moon \citep{lissauer1997s}.

From an observational point of view, in \cite{borucki1984photometric} is shown that   eclipsing binaries are good targets for detecting planets through transit searches. It is shown that eclipsing binaries have three advantages in detecting planets. (1) The orbital plane of the planetary system must be close to the line of sight, because is expected that the angular moment vector of the binary pair rotating around the common center of mass between them is greater than in systems with only one star with equal mass of just one of the binary components. (2) Each one of the stars may have planets, considering systems where the stars are wide separated, and (3) planets orbiting close-binary stars, will have a shorter orbital period than if they were orbiting a star of the same spectral class. In addition they will have two stars to transit instead of one, giving a lot more information.  
In the case of CB systems that already have a planet in transit (which is the core of our present work), \cite{kratter2014planet} showed that this configuration is a strong indicator that other planets must also be in transit. With that, knowing the possibility of
a CB Earth-type planets being formed and their locations in real systems with planets already detected, can contribute to the detection of the first planet of this type in a binary system and with the possibility of having life.

Therefore, in this work we are looking for real CB systems with confirmed planets, Earth-like conditions and consequently Earth-like life. Then, in the next section we checked from among all confirmed circumbinary planets, which ones have enough data needed for our study. With the selected systems, we computed their habitable zone. In section \ref{sec:3} we checked the stability of massless particles within the limits of the habitable zones. In section \ref{sec:4},  we used numerical simulations to explore the formation of terrestrial planets within the stable regions 
determined in section \ref{sec:3}. In the last section we make our final remarks.

\section{Binary stars systems and their HZ}
\label{sec:2}
\begin{table}
\centering
\begin{tabular}{lcc}
\hline
\hline
System            & Discovery year  & Number of Planets\\
\hline
DP Leo            & 2009    & 1  \\
FW Tauri          & 2013    & 1  \\
HD 106906         & 2013    & 1  \\
HU Aqr            & 2011    & 1  \\
Kepler-16         & 2011    & 1   \\
Kepler-34         & 2012    & 1    \\
Kepler-35         & 2012    & 1   \\
Kepler-38         & 2012    & 1    \\
Kepler-47         & 2012 - 2019    & 3   \\
Kepler-413        & 2013    & 1    \\
Kepler-451        & 2015    & 1   \\
Kepler-453        & 2014    & 1    \\
Kepler-1647       & 2015    & 1   \\
NN Ser AB         & 2010    & 2   \\
NY Vir AB         & 2011    & 2  \\
OGLE-2007-BLG-349 & 2016    & 1  \\
PSR B1620-26      & 2003    & 1  \\
Ross 458          & 2010    & 1   \\
Roxs 42           & 2013    & 1   \\
RR Cae            & 2012    & 1   \\
SR 12 AB          & 2011    & 1    \\
TOI-1338 AB       & 2020    & 1    \\
\hline
\end{tabular}
\caption{Close Binary stars systems with confirmed planets \protect{\citep{schwarz2016new}}. 
The \textit{Kepler-47} system has two discovery dates, 2012 and 2019, because two planets in the system were confirmed in 2012 and another in 2019.}
\label{systems}
\end{table}

As we are exploring the formation of planets within the HZ of their systems, we first select all binary systems that have planets to compute their HZ, see Table \ref{systems}. However, not all systems chosen had the data required for this calculation or were not stars of the main sequence, such as the \textit{PSR B1620-26} system which is a pulsar. Only some of the binary systems detected by the Kepler probe remain (\textit{Kepler-16, 34, 35, 38, 47, 413, 453} and \textit{1647}). The data of these systems are given in Table \ref{parameters_systems}. For the selected systems, we computed the boundaries of their respective HZ using two different models present in \cite{haghighipour2013calculating} and \cite{mason2015circumbinary}, see Table \ref{hz_os_systems}. We can note that boundaries of the HZ of each system are very similar using the two models. So, in our plots and analyzes we will take into account only the model of \cite{haghighipour2013calculating}.

  \begin{table*}
	\centering
	\begin{tabular}{lccccccccc}
		\hline	    
		\hline	    
		$System$ & $M_{A}$ ($M_{\odot}$) & $M_{B}$ ($M_{\odot}$) & $a_{bin}$(au) & $e_{bin}$ & $a_{p}$(au) & $e_p$ & $M_p$ ($M_j$) & $i_p$ (Deg) & References\\
		\hline	
		K-16 & 0.689 & 0.203 & 0.224 & 0.159 & 0.705 & 0.007 & 0.333 & 0.308&\cite{doyle2011kepler}\\		 
		K-34 & 1.048 & 1.021 & 0.229 & 0.521 & 1.089 & 0.182 & 0.220& 0.497&\cite{welsh2012transiting}\\
		K-35 & 0.890 & 0.810 & 0.176 & 0.142 & 0.603 & 0.042 & 0.127& 0.336&\cite{welsh2012transiting}\\
		K-38 & 0.949 & 0.249 & 0.147 & 0.103 & 0.464 & 0.030 & 0.016& 0.182&\cite{orosz2012neptune}\\
		K-47 & 1.043 & 0.362 & 0.084 & 0.023 & 0.297 & 0.035 & 0.028 & 0.270&\cite{orosz2012kepler}\\
		     &       &       &       &       & 0.989 & 0.411 & 0.061 & 1.160&\cite{orosz2012kepler}\\
		     &       &       &       &       &      0.699  & 0.024       & 0.060     &     0.000 & \cite{orosz2019discovery}\\
		K-413 & 0.820 & 0.542 & 0.102 & 0.037 & 0.355 & 0.118 & 0.210& 4.073&\cite{kostov2014kepler}\\
		K-453 & 0.944 & 0.195 & 0.185 & 0.052 & 0.790 & 0.036 & 0.030& 2.258&\cite{welsh2015kepler}\\
		K-1647 & 1.221 & 0.968 & 0.128 & 0.159 & 2.721 & 0.0581 & 1.520 & 2.986 &\cite{kostov2016kepler}\\
		\hline	 	   
	\end{tabular}
	\caption{The main parameters of the eight selected binary systems with the potential to form Earth-like planets in their HZ. $M_A$ and $M_B$, $a_{bin}$ and $e_{bin}$ are the masses, semi-major axis and eccentricity from the binary pair respectively and $a_{p}$, $e_p$, $M_p$ e $i_p$ are semi-major axis, eccentricity, mass and inclination with respect to with the binary plane of the planet of each system.}             
	\label{parameters_systems}
\end{table*}

\begin{table*}
\newcommand{\mc}[3]{\multicolumn{#1}{#2}{#3}}
\begin{center}

\begin{tabular}{lclllcclll}
\hline
\hline

& \mc{4}{c}{Haghighipour \& Kaltenegger}& & \mc{4}{c}{Mason et al.}\\
\cline{2-10}
& \mc{2}{c}{Narrow HZ} & \mc{2}{c}{Wide HZ}& & \mc{2}{c}{Narrow HZ} & \mc{2}{c}{Wide HZ}\\
\cline{2-5}
\cline{7-10}
\multirow{-3}{*}{Systems} & Inner (au) & \mc{1}{c}{Outer (au)} & \mc{1}{c}{Inner (au)} & \mc{1}{c}{Outer (au)} && Inner (au) & \mc{1}{c}{Outer (au)} & \mc{1}{c}{Inner (au)} & \mc{1}{c}{Outer (au)}\\
\hline
Kepler-16 & 0.39 & \mc{1}{c}{0.65} & \mc{1}{c}{0.30} & \mc{1}{c}{0.70} && 0.45 & \mc{1}{c}{0.82} & \mc{1}{c}{0.36} & \mc{1}{c}{0.87}\\
Kepler-34 & 1.55 & \mc{1}{c}{2.73} & \mc{1}{c}{1.20} & \mc{1}{c}{2.90} && 1.62 & \mc{1}{c}{2.83} & \mc{1}{c}{1.28} & \mc{1}{c}{2.98}\\
Kepler-35 & 1.10 & \mc{1}{c}{1.98} & \mc{1}{c}{0.89} & \mc{1}{c}{2.09} && 1.18 & \mc{1}{c}{2.07} & \mc{1}{c}{0.93} & \mc{1}{c}{2.18}\\
Kepler-38 & 1.60 & \mc{1}{c}{2.85} & \mc{1}{c}{1.30} & \mc{1}{c}{3.00} && 0.91 & \mc{1}{c}{1.61} & \mc{1}{c}{0.72} & \mc{1}{c}{1.61}\\
Kepler-47 & 0.89 & \mc{1}{c}{1.58} & \mc{1}{c}{0.69} & \mc{1}{c}{1.65} && 0.89 & \mc{1}{c}{1.57} & \mc{1}{c}{0.70} & \mc{1}{c}{1.66}\\
Kepler-413 & 0.50 & \mc{1}{c}{0.85} & \mc{1}{c}{0.40} & \mc{1}{c}{0.88} && 0.57 & \mc{1}{c}{1.04} & \mc{1}{c}{0.45} & \mc{1}{c}{1.10}\\
Kepler-453 & 0.70 & \mc{1}{c}{1.25} & \mc{1}{c}{0.55} & \mc{1}{c}{1.31} && 0.75 & \mc{1}{c}{1.33} & \mc{1}{c}{0.59} & \mc{1}{c}{1.41}\\
Kepler-1647 & 2.10 & \mc{1}{c}{3.79} & \mc{1}{c}{1.57} & \mc{1}{c}{4.00} && 2.34 & \mc{1}{c}{4.08} & \mc{1}{c}{1.85} & \mc{1}{c}{4.30}\\
\hline
\end{tabular}
\caption{Selected close binary stars systems with their respectively HZ using two models to determine their bounds.}
\label{hz_os_systems}
\end{center}
\end{table*}

\section{Stability inside the Habitable Zone}
\label{sec:3}
With known HZ, since our objective is to study the possibility of formation of terrestrial planets within these regions, we first check if they are stable or at least have stable parts. The stability study will be important because the results of these tests will be the initial conditions for the formation simulations.
Using an adaptation made by us (called \textit{Minor-Mercury package} \citep{Amarante2019}) of the numerical integrator of N-body \textit{Mercury} \citep{chambers1999hybrid} based on \cite{chambers2002symplectic}, with the hybrid integrator option for binary systems, we performed numerical simulations with test particles (particles that do not interact gravitationally with each other), the binary pair and the host planets of each system. 

\subsection{Initial conditions of the test particles}

The parameters of the eight systems selected are the ones shown in Table \ref{parameters_systems}. The particles were distributed randomly along the HZ with an extra margin of 20\% of the total width of the HZ for each side, i.e. internal and external, with eccentricity equal to zero and coplanar with the orbital plane of the binaries. The number of particles for each system varied according to the width of each region to be distributed. For this, we use a density of particles per $au^2$ of $\approx$ 160. Particles were removed from the simulation if their semi-major axis exceeded 10 $au$. Thus, eight simulations were performed with integration final time of 1 $Myr$.

   \begin{figure*}
   \centering
   \includegraphics*[scale=0.55]{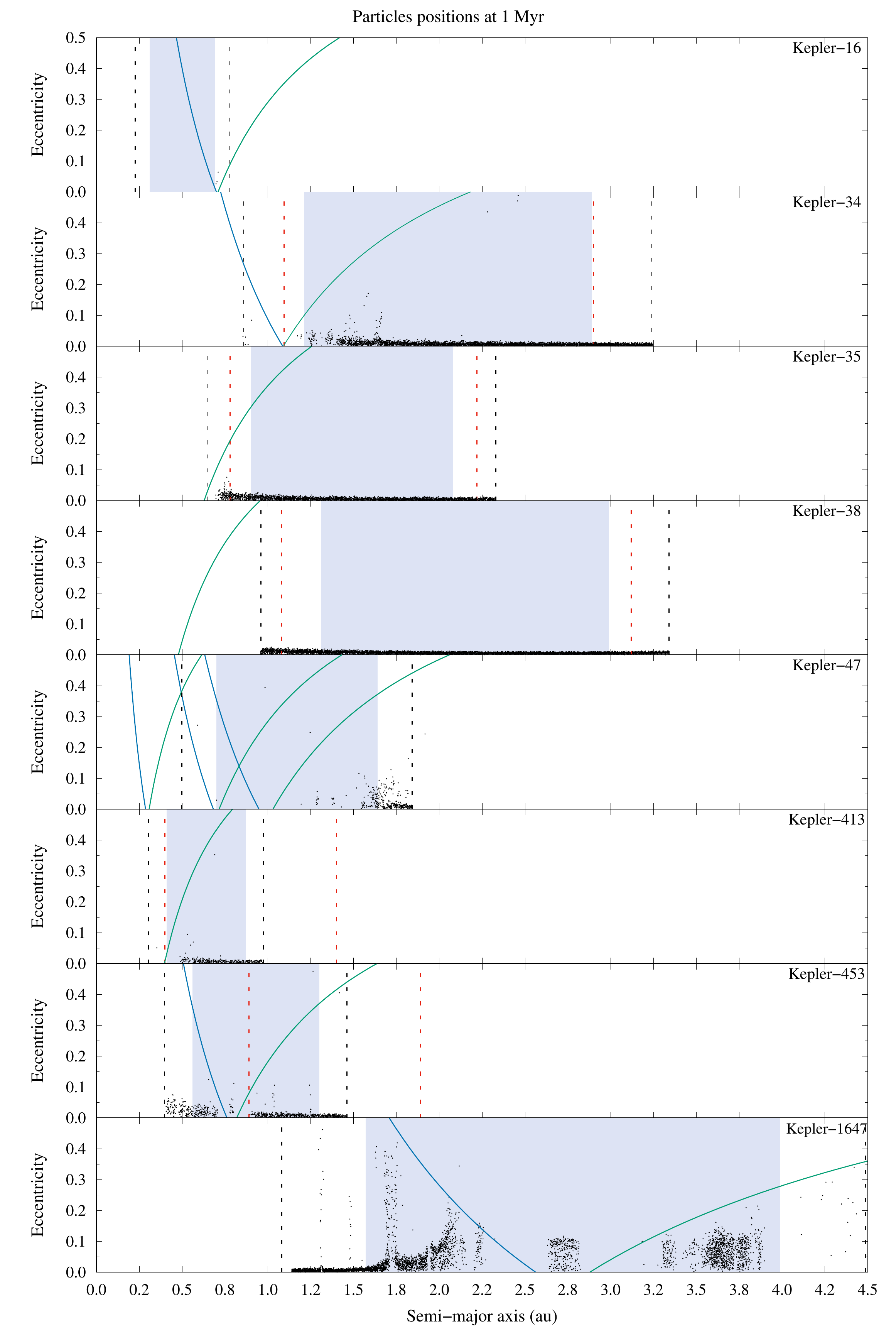}
      \caption{Final positions of the test particles for the selected systems (\textit{Kepler-16, 34, 35, 38, 47, 413, 453} and \textit{1647}). The region shaded in blue indicates the habitable zone of each system. The green lines in each subfigure represent the apocenter of the host planets of each system as a function of the pericenter of the particles given by $a=[a_p(1+e_p)]/(1-e)$ and the blue lines represent the pericenter of the planet as a function of the apocenter of the particles given by $a=[a_p(1-e_p)]/(1+e)$. $a_p$ and $e_p$ represent the semi-major axis and eccentricity of the host planet respectively and $a$ and $e$ are the same for the particles. The black dashed lines indicate the limits of the initial distribution of the test particles and the red dashed lines indicate the limits of the stable region to be used for the distribution of bodies to the planetary formation study.}
         \label{stability_final}
  \end{figure*}

The internal and external limits for the HZ of each system can be checked in Table \ref{hz_os_systems}. We consider the limits of empirical (wider) HZ to distribute the particles. The extra limits where the particles were distributed can be checked in the Figure \ref{stability_final}, where they are represented by dashed black lines. 

\subsection{Results}

\subsubsection{Kepler-16}

The sum of the masses of the binary pair equals approximately 0.9 $M_\odot$, so it is the least massive pair in our simulations. Thus, it is the system that has the narrowest and closest HZ to the system's barycenter, causing much disturbance in the particle disk by the motion of the stars. Another factor that increases disk disturbance is the semi-major axis of the stars, which is the second largest of our test cases, equal to 0.224 $au$. Also, the system planet is inside of HZ. These ingredients cause all particles to be ejected within a few thousand years. Only some coorbitant particles to the host planet survived in the simulation, see Figure \ref{stability_final}. Which indicates that the habitable region of the system is unstable. This result shows us that forming a planet in HZ is difficult, but this does not exclude the possibility of any body orbiting these regions. Since this system has unstable HZ, we do not simulate the formation of planets in this case.

Although this system is unstable throughout its HZ, it is possible to note that the system has bodies coorbital to the host planet, see Figure \ref{stability_final}. This is an important result mainly for two reasons; (1) demonstrates the condition of binary systems having bodies coorbital to the host planets; (2) As in this case the planet is within the HZ, consequently coorbital bodies will also be. Thus, even if the HZ is dynamically unstable, the system can still have bodies within the HZ sharing the orbit with the host planet.

\subsubsection{Kepler-34}

The system's HZ is away from the system center of mass. This makes the disk not very disturbed by the stars. The dominant disk disturbance comes from the confirmed planet. The planet is positioned at the beginning of HZ and consequently on the particle disk, see Figure \ref{stability_final}. This causes further disturbance in the early part of the disk. Even with the planet's presence, HZ is practically entirely stable. Therefore, the system presents stability conditions to study the formation of planets in HZ.  

\subsubsection{Kepler-35}

The \textit{Kepler-35} system is very similar with \textit{Kepler-34}. The system also has the planet very close to the interior of the habitable region, which makes the planet the main disturbing agent of the disk. Although close, the planet never orbits within to HZ as in the previous case, so what makes it completely stable and the particles practically did not vary their positions. This can be checked in Figure \ref{k35_stability_4}, where snapshots of the evolution of the semimajor axis and eccentricity in the \textit{Kepler-35} system are shown. Only the particles closest to the planet were ejected giving us a large region for the study of planetary formation. 

\subsubsection{Kepler-38}

\begin{figure}
   \centering
   \includegraphics*[width=\hsize]{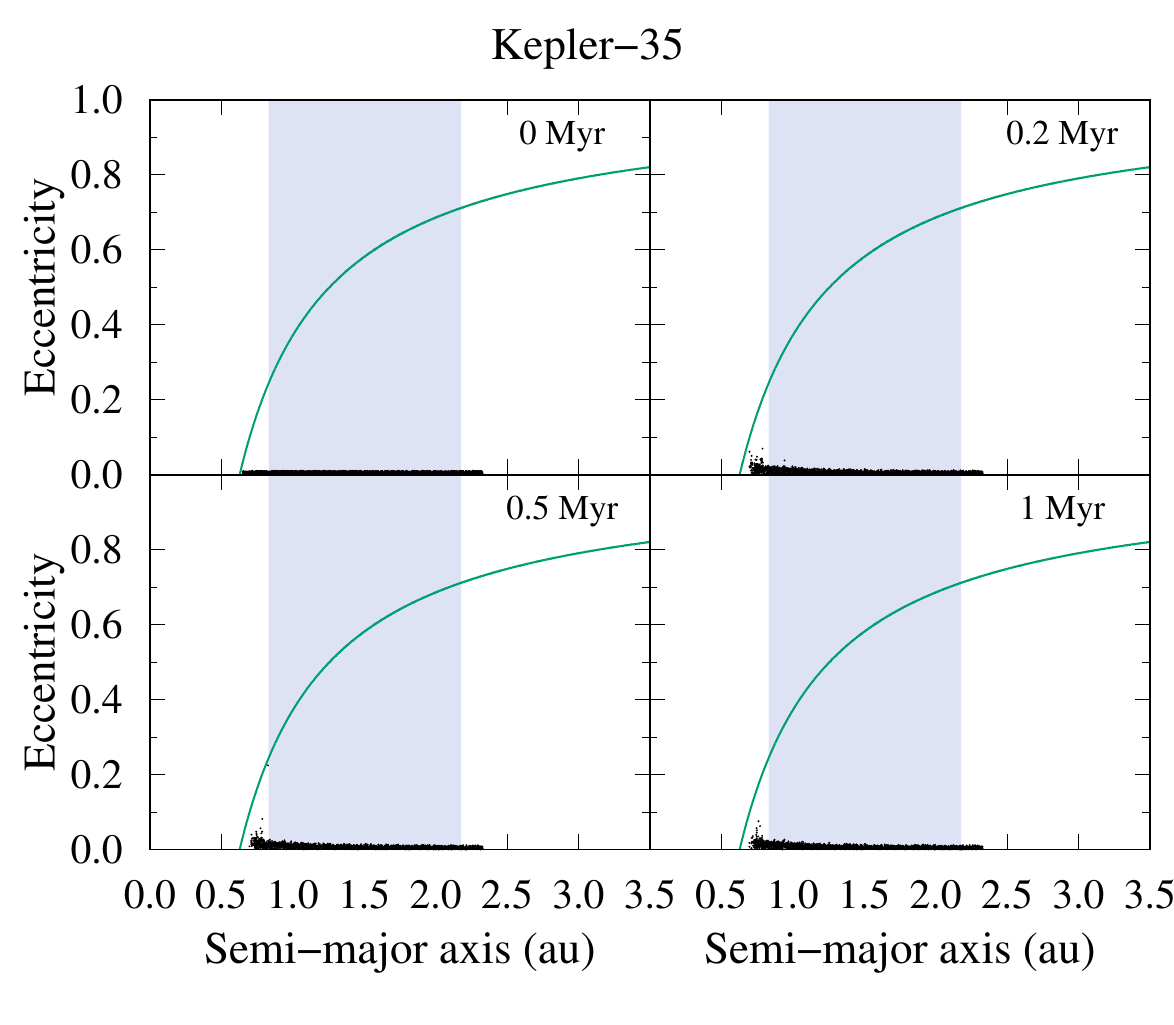}
      \caption{Snapshots in time of the dynamic evolution of particles in the \textit{Kepler-35} system. The horizontal and vertical axes are the semi-major axis and the eccentricity, respectively.The black dots are the particles and the green line is the apocenter of the host planet of the system in function of the pericenter of the particle given by $a=[a_p(1 + e_p)]/(1-e)$. The shaded region in blue represents the HZ of the system.}
         \label{k35_stability_4}
   \end{figure}

This system is the most stable of all studied in this work, as can be seen in Figure \ref{k35_stability_4}. This feature comes from the great distance that HZ is from the planet and consequently from the stars. In addition, the planet is the least massive of all eight selected. with reasonable mass of 0.016 Jupiter masses. Throughout the simulation, no particles were ejected, and as we can see from Figure \ref{k35_stability_4}, only the early disk particles had a slight increase in eccentricity. 
Thus, the system has a large region for the study of planetary formation within its HZ.

\subsubsection{Kepler-47}

The \textit{Kepler-47} system has already been the subject of a stability study in some works before the third planet was confirmed. In \cite{kratter2014planet}, the system was used as a case study to verify that the system is dynamically packed with the three planets. In \cite{hinse2015predicting}, a stability study was performed using MEGNO technique to predict the third planet.

\begin{figure}
   \centering
   \includegraphics*[width=\hsize]{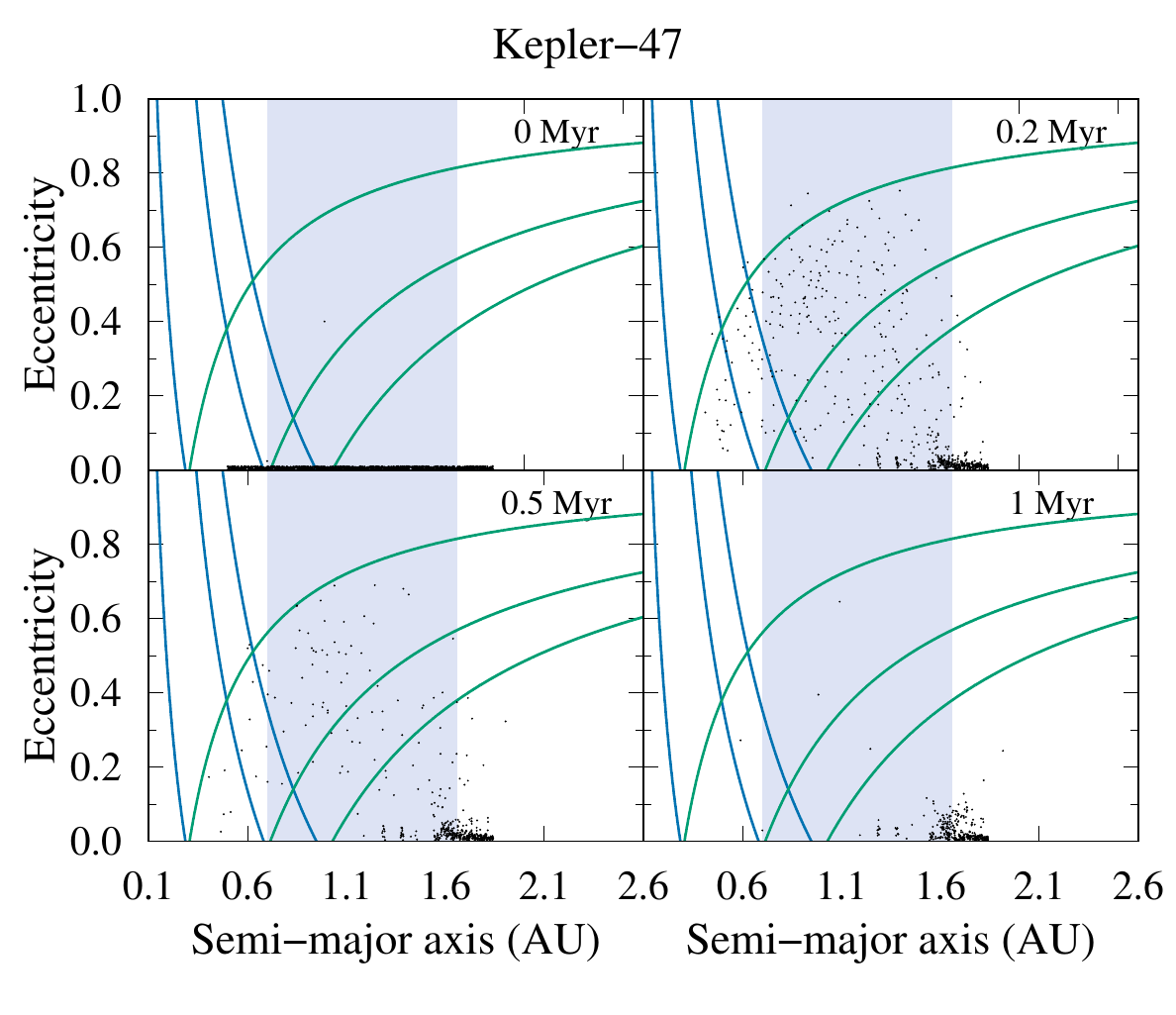}
      \caption{Snapshots in time of the dynamic evolution of particles in the \textit{Kepler-47} system. The horizontal and vertical axes are the semi-major axis and the eccentricity, respectively. The black dots are the particles and green lines represent the apocenter of the host planets of the system in function of the pericenter of the particle, given by $a=[a_p(1 + e_p)]/(1-e)$ and the blue lines represent the pericenter of the planet as a function of the apocenter of the particles given by $a=[a_p(1 - e_p )]/(1 + e)$. The shaded region in blue represents the HZ of the system.}
         \label{k47_stability_4}
   \end{figure}

In our work, with the confirmation of the three planets, we used the three to test the dynamic stability of the HZ. Thus, this case is the only multiplanetary system present in our stability test. The most recent confirmation is from the planet \textit{Kepler-47d} \citep{orosz2019discovery}. One of the planets is orbiting within the HZ, which causes great instability in that region. We can check in Figures \ref{stability_final} and \ref{k47_stability_4} that almost all particles were ejected at 1 $Myr$, with only a few particles remaining on the outside of the HZ. Thus, we can conclude that HZ is unstable to its full extent and does not provide conditions to study the ability of this system to form a planet within HZ.

Together with the \textit{Kepler-16} system, this case is the most unstable HZ of our simulations. In addition to the instability present in both cases, one of the host planets also has a coorbital body to it, see Figure \ref{stability_final}.

 \subsubsection{Kepler-413}
 
 This system has a very narrow HZ and a host planet on its inner edge.
 In this case, the particles that are in the innermost region of the disk are ejected because they are very close to the planet. Although many particles have been ejected and collided with the planet, a reasonable fraction of the HZ of the \textit{Kepler-413} system has stability. 
 
  \subsubsection{Kepler-453}
 
 The planet of the system orbits within to HZ, which causes a lot of ejections and collisions of the particles. We can see in Figure \ref{stability_final} two regions, one at the beginning and one at the end of HZ, that have stability, the largest being the outermost region. So even though it has no stability to its full extent, the system has parts that are stable.
 
Besides the system has a stable part of the HZ, the same result of the \textit{Kepler-16} system can be noticed in this case, see Figure \ref{stability_final}. The system has bodies coorbital to the host planet as well.
 
  \subsubsection{Kepler-1647}
  
The \textit{Kepler-1647} system has the largest exoplanet in binary systems with approximately 1.5 Jupiter masses. In addition to the planet, stars also have the largest mass sum in our work, which makes HZ farther and wider. The planet is in the middle of HZ, causing many particles to be ejected. In Figure \ref{stability_final} we can see three subregions within the HZ. An inner region, an coorbital region and another outer region in relation to the planet. In this case, because it is a peculiar and complex system, we will study the formation of terrestrial planets within these three regions in a future work.

With the stable regions found in the above systems, we can now study the ability to form terrestrial planets in these regions. Some systems, as in the case of \textit{Kepler-16, 47} and \textit{1647}, do not have a stable region sufficient for this study. Others, such as the \textit{Kepler-413} and \textit{453} systems, do not have completely stable HZ, but we can study at least some parts of these regions. The red dashed lines in Figure \ref{stability_final} are the inner and outer limits where we distribute the embryos and planetesimals to explore the terrestrial planet formation.

\section{PLANETARY FORMATION}
\label{sec:4}
Because our work is to explore terrestrial planet formation around binary stars systems, we followed the classical paradigm of terrestrial planet formation in single star systems because theories about the stages of planetary formation and consequently, formation of terrestrial planets in binary systems do not yet exist. Our simulations are performed starting with two different surface density profiles given by $\Sigma_1r^{-x}$ where $x=1.5$ and $2.5$. $\Sigma_1$ is a solid surface density at 1 $au$. The $\Sigma_1$ was ajusted to fix the total mass of 2.5$M_\oplus$ \citep{quintana2004planet} (sum of the masses of the terrestrial planets in solar system) in the stable and habitable region. 

\begin{figure*}
\centering
        \begin{subfigure}{0.335\textwidth}
                \includegraphics*[width=\linewidth]{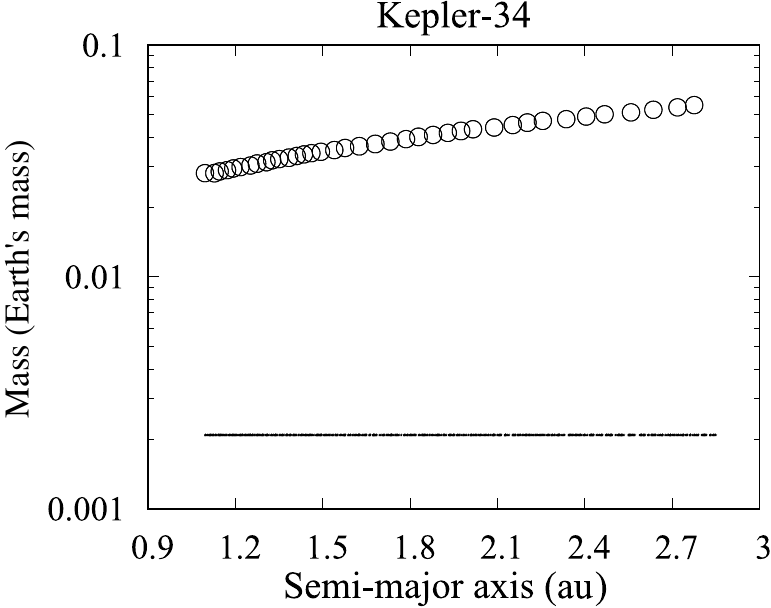}
        \end{subfigure}%
        \begin{subfigure}{0.335\textwidth}
                \includegraphics*[width=\linewidth]{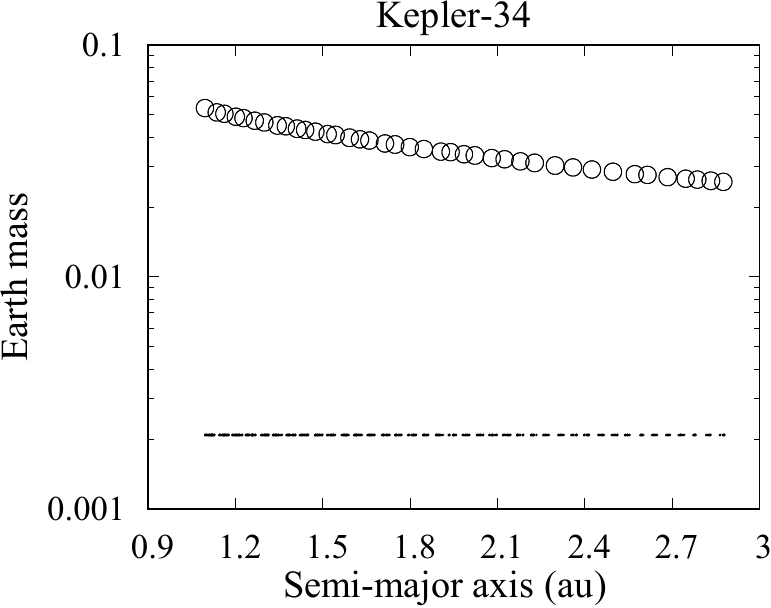}
        \end{subfigure}%
        \\
        
        \begin{subfigure}{0.338\textwidth}
                \includegraphics*[width=\linewidth]{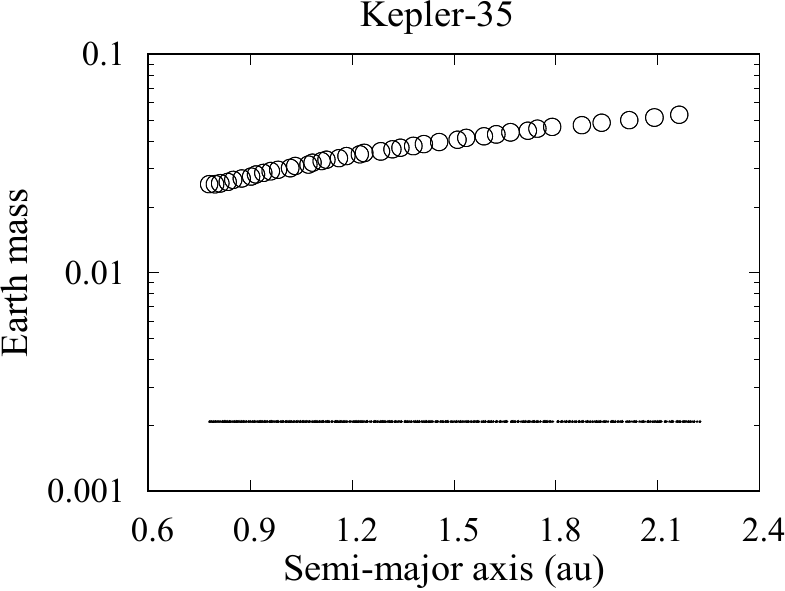}
        \end{subfigure}%
        \begin{subfigure}{0.338\textwidth}
                \includegraphics*[width=\linewidth]{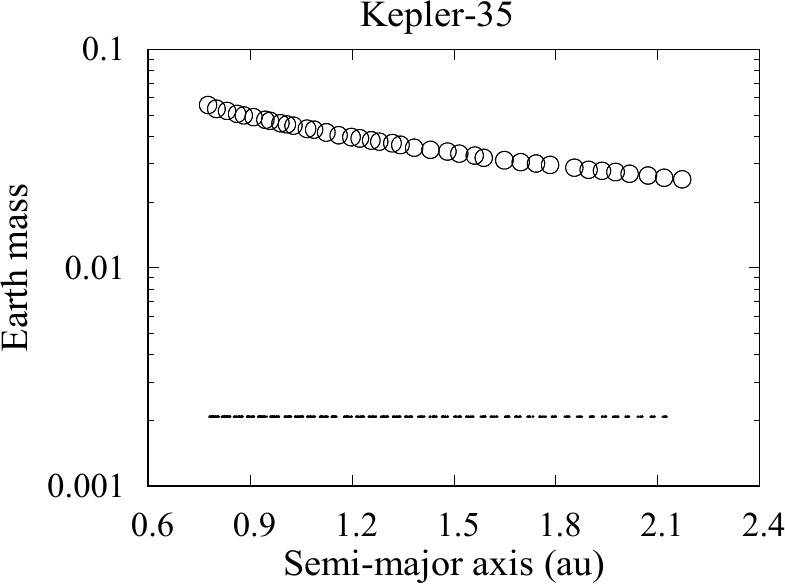}
        \end{subfigure}%
        \\
        \begin{subfigure}{0.33\textwidth}
                \includegraphics*[width=\linewidth]{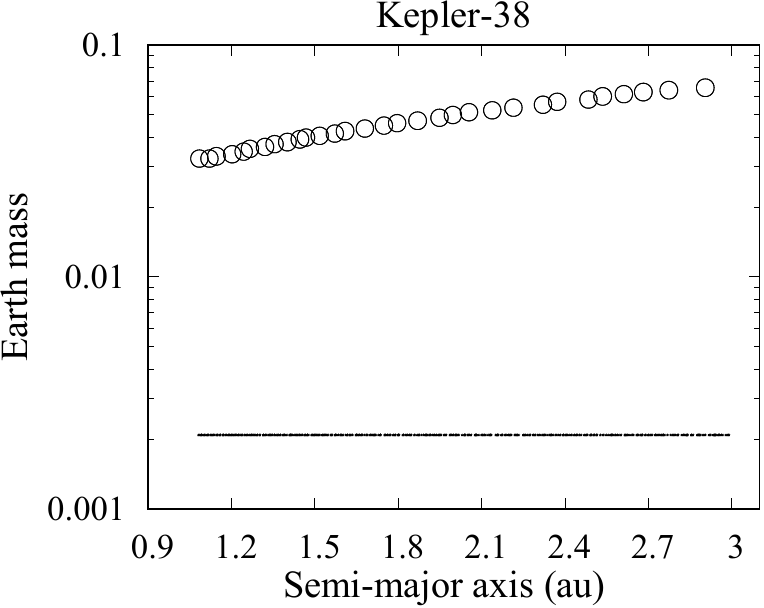}
        \end{subfigure}%
        \begin{subfigure}{0.33\textwidth}
                \includegraphics*[width=\linewidth]{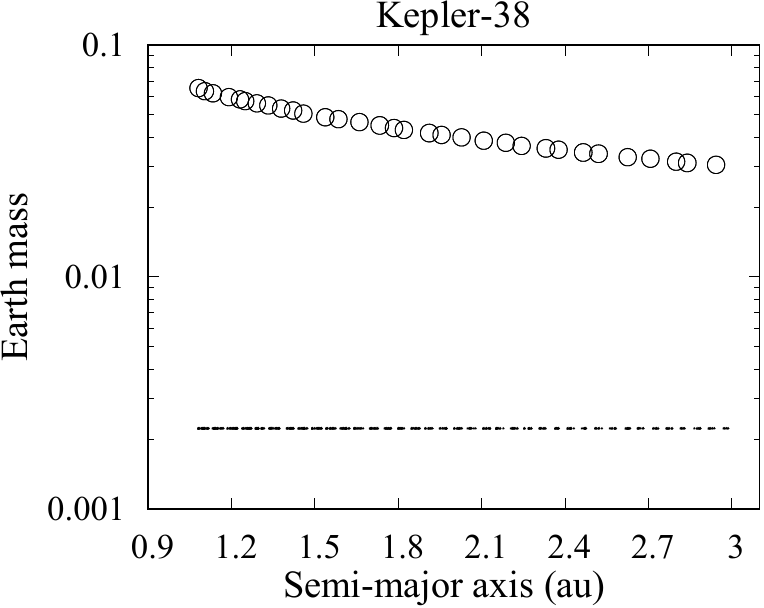}
        \end{subfigure}%
        \\
        \begin{subfigure}{0.33\textwidth}
                \includegraphics*[width=\linewidth]{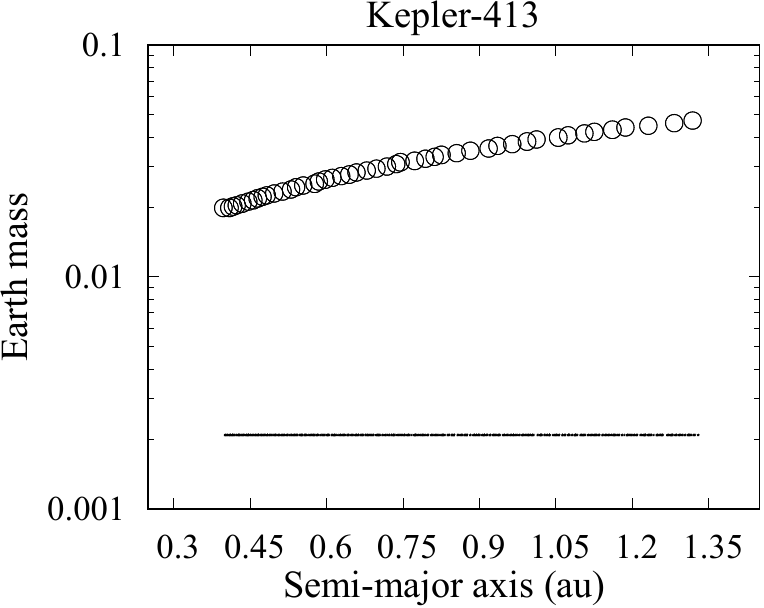}
        \end{subfigure}%
         \begin{subfigure}{0.33\textwidth}
                \includegraphics*[width=\linewidth]{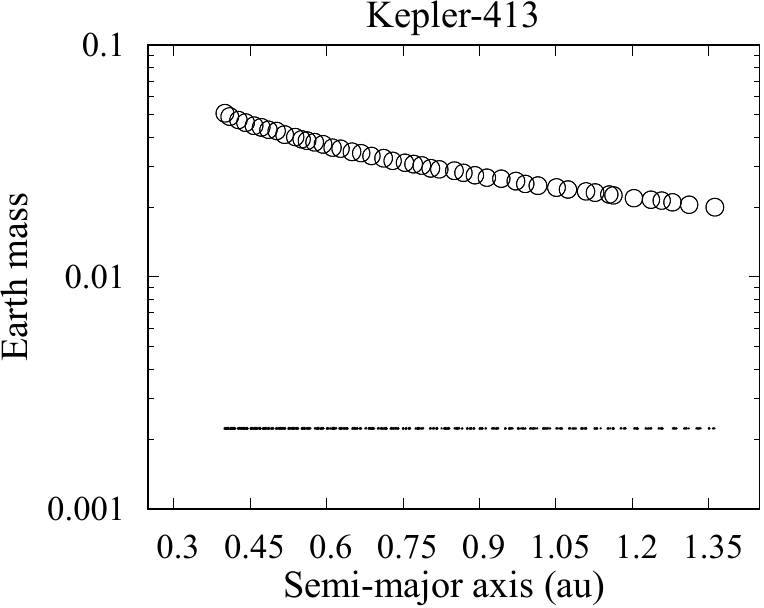}
        \end{subfigure}%
        \\
        \begin{subfigure}{0.33\textwidth}
                \includegraphics*[width=\linewidth]{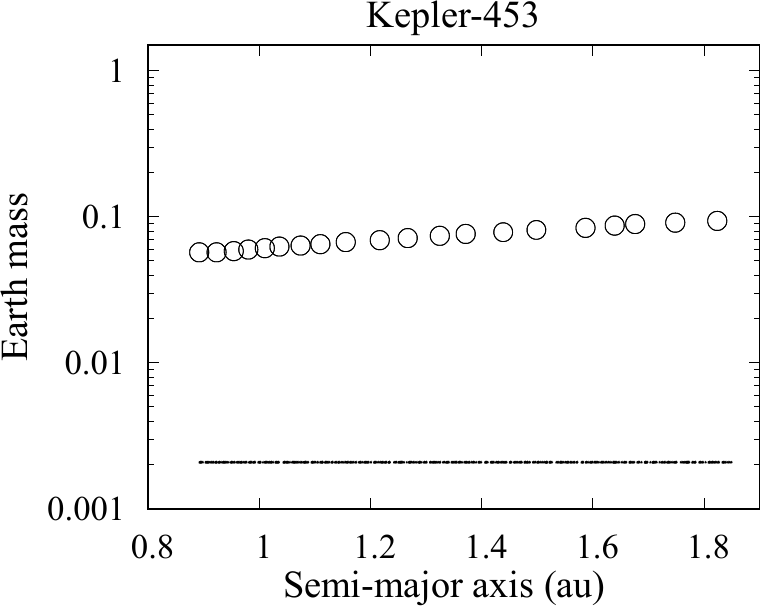}
        \end{subfigure}%
        \begin{subfigure}{0.33\textwidth}
                \includegraphics*[width=\linewidth]{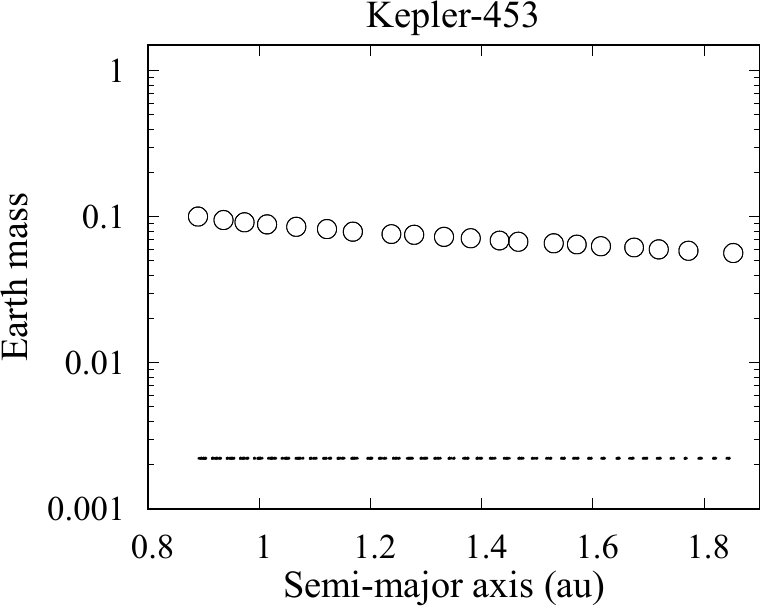}
        \end{subfigure}%

\caption{Initial conditions for the selected binary systems. In the left and right columns are the initial distributions of planetesimals and embryos for $x$ = 1.5 and 2.5, respectively.}
\label{init_1}
\end{figure*}

In our simulations, we assumed that the growth of dust grains and planetesimals occurred during the first stages of planetary accretion. Thus, we used a bimodal disk composed of embryos (60\% of the disk's mass) and planetesimals (40\% of the disk's mass) \citep{izidoro2015terrestrial}. It was assumed that embryos are formed by oligarchic growth and are thus spaced randomly by 5-10 mutual Hill radii \citep{kokubo1998oligarchic, kokubo2000formation} with a density of 3$g/cm^3$. The mass of each planetesimals is $\approx$ 0.002$M_\oplus$. In the numerical integrations, the planetesimals do not have gravitational interactions with themselves, only with stars, planets and protoplanetary embryos. The protoplanetary embryos masses scale as $M\approx r^{3(2-x)/2}\Delta^{3/2}$ \citep{kokubo2002formation, raymond2005terrestrial, raymond2009building, izidoro2015terrestrial} where $\Delta$ is the mutual Hill radii separations between embryos orbits. As we are using distinct systems with different parameters, the number of embryos and planetesimals are not the same among the systems, see Table \ref{number_bodys}. Figure \ref{init_1} shows the initial conditions for $x=1.5$ and $2.5$, for all our simulations. Disks with $x=2.5$ have embryos more massives in the inner region of the disk, and with 1.5 in the outer region. We chose these two values of the parameter $x$, besides being used frequently in some works \citep{raymond2004making, raymond2005terrestrial, izidoro2014terrestrial, izidoro2014terrestrial2, izidoro2015terrestrial, izidoro2018formation}, to study the dynamic evolution in disks that have mass growing as a function of the orbital radius of the bodies and in cases where the mass decreases as a function of the orbital radius (see Figure \ref{init_1}). This is an important point because we have some systems where the giant planet is in the inner region of the disk and others in the outer region. So we needed to consider the two cases for all systems. The orbital inclination of the embryos and planetesimals are chosen randomly between $10^{-4}$$^{\circ}$ and $10^{-3}$$^{\circ}$ with respect to the binary plane, and the eccentricity are chosen between 0 and 0.01.

In our simulations were used the parameters from the stars and planets hosts of each system given in Table \ref{parameters_systems}. For each surface density profile were performed 5 simulations with a few differences generated randomly for the initial disk, totalizing 50 simulations. Thus were integrated for 200 $Myr$ \citep{quintana2004planet} using an adaptation of the \textit{Mercury} package \citep{chambers1999hybrid}, made by us following the work \cite{chambers2002symplectic} as previously mentioned. Collisions between planetary embryos and with planetesimals are considered inelastic such that in each collision mass and linear momentum are conserved and a new body is formed.

\begin{table}
{
\centering
\subcaption*{$x$ = 1.5}
\begin{tabular}{lcc}
\hline
\hline
Systems & Number of Planetesimals & Number of Embryos \\
\hline
Kepler-34  & 480                     &  40                \\
Kepler-35  & 480                     &  40                \\
Kepler-38  & 480                     &  35                \\
Kepler-413 & 470                     &  48                \\
Kepler-453 & 450                     &  20                \\
\hline
\end{tabular}
\centering
\subcaption*{$x$ = 2.5}
\begin{tabular}{lcc}
\hline
\hline
Systems    & Number of Planetesimals  & Number of Embryos  \\
\hline
Kepler-34  & 475                     &  40                \\
Kepler-35  & 470                     &  40                \\
Kepler-38  & 440                     &  35                \\
Kepler-413 & 470                     &  47                \\
Kepler-453 & 440                     &  20                \\
\hline
\end{tabular}

}
\caption{The approximate number of planetesimals and embryos of each system for
two different value of $x$ (1.5 and 2.5).}
\label{number_bodys}

\end{table}

\subsection{Results and discussion}

\begin{figure*}
   \centering
   \includegraphics*[scale=0.53]{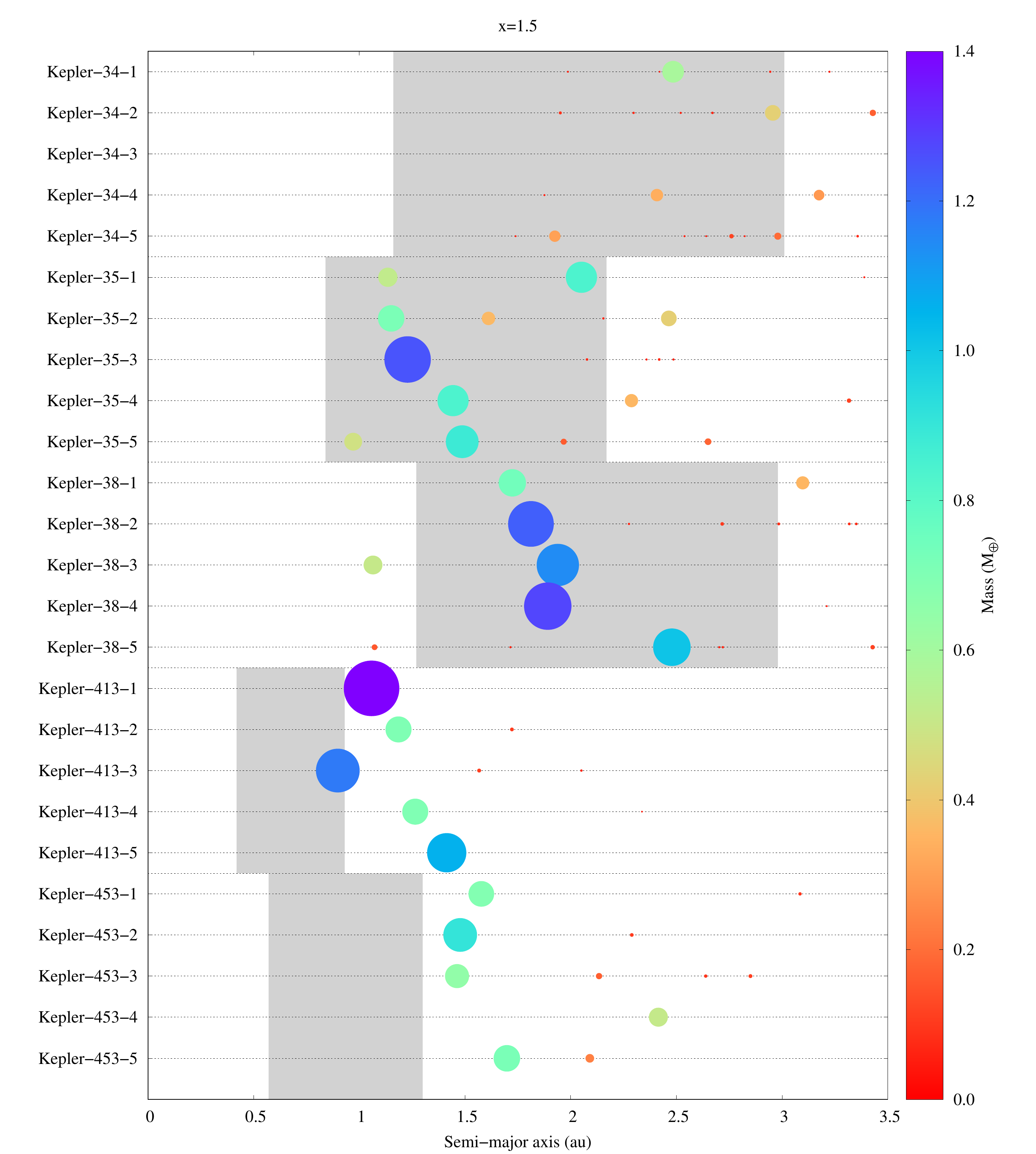}
      \caption{Final orbital configuration of 25 simulations of the systems using $x=1.5$. The circles represent the bodies formed in the simulations and their sizes are proportional to their masses. The color of the plants represent their mass in terms of Earth masses, and the shaded regions are the HZs of each system.}
         \label{1}
  \end{figure*}

\begin{figure*}
   \centering
   \includegraphics*[scale=0.53]{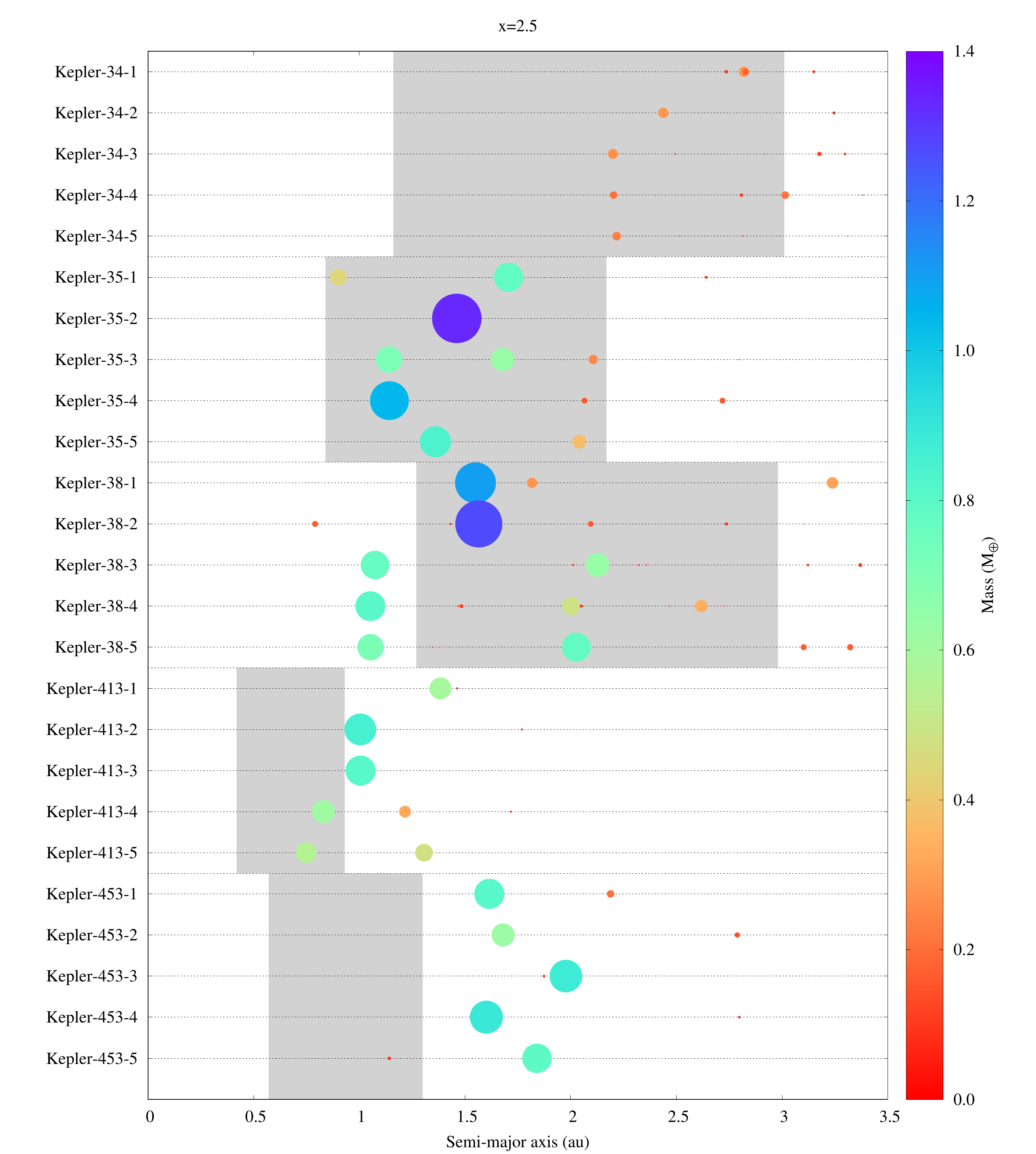}
      \caption{Final orbital configuration of 25 simulations of the systems using $x=2.5$. The circles represent the bodies formed in the simulations and their sizes are proportional to their masses. The color of the plants represent their mass in terms of Earth masses, and the shaded regions are the HZs of each system},
         \label{2}
   \end{figure*}

In this section we will see the individual results of the numerical simulations of each system, addressing the different values of the coefficient \textit{x} (1.5 and 2.5). 
It is important to remember that $x=1.5$ results in a disk in which the mass of the embryos is increasing according to the semi-major axis of the embryo, and values of $x=2.5$, provide disks with decreasing embryo mass along the disk (Figure \ref{init_1}). The results of our numerical simulations, referring to all systems, are present in Figures \ref{1} and \ref{2}, referring to the final integration time with $x = 1.5$ and $x = 2.5$, respectively. In these figures are shown the semi-major axis of the planets formed in five simulations for each system. The size of each circle in the plots is proportional to the mass of the planet generated in the simulation and the colors provide their mass with respect to the mass of the Earth. 
As the limits of the HZ vary with the mass of the planet \citep{kopparapu2014habitable}, we consider as an Earth-size a planet that has mass varying between 0.80 - 1.20 M $_\oplus$.

\subsubsection{Kepler-34}

The \textit{Kepler-34} system has two stars with masses close to the Sun (see parameters in Table 3), consequently belong to spectral class G with age of 5-6 \textit{Gyr}. They complete an orbital period in about 28 days \citep{welsh2012transiting}. This system has one giant planet with $\approx$ 22\% of the mass of Jupiter, 76\% of the radius of Jupiter and is known
as \textit{Kepler-34b}. The \textit{Kepler-34} system has its planet very close to the disk of planetesimals and embryos. This makes the planet the main disturber. This causes a lot of disk mass to be lost as early as the first thousand of years of integration, especially in the early part of the disk. At $ x = 1.5 $, 24\% of the initial mass was lost, while in the case of $x = 2.5$ 36\% of the mass was lost at the same length of time. The percentage is higher in the case of $ x = 2.5 $, because in this case the most massive embryos are at the inner part of the disk, the most disturbed region. Thus being ejected or collided with the planet. With both mass scale parameters, protoplanets are formed within the HZ in all simulations. The closest to a terrestrial mass appears in the case \textit{Kepler-34-1} (with $x=1.5$), where a planet with a mass of about 0.6 $M_{\oplus}$ is formed in the simulation, see Figure \ref{1}. In all others, with both values of $x$, only bodies with masses < 0.4 $M_{\oplus}$ are formed, see Figures \ref{1} and \ref{2}. Therefore, from our initial conditions, this system can form a terrestrial planet with up to 0.6 Earth mass within HZ.

\subsubsection{Kepler-35}

This system also has stars with mass close to that of the Sun, and with this also belonging to the spectral class G. These stars have a 0.176 $au$ semi-major axis and an orbital period of 21 days with an age of 8-12 Gyr \citep{welsh2012transiting}. The host planet of the system, known as \textit{Kepler-35b}, has 13\% of the mass and 73\% of the radius of Jupiter \citep{welsh2012transiting}, it is $\approx$ 0.3 $au$ away from the beginning of the disk. In the stability test within the HZ, we saw that the test particles hardly changed in their semi-major axis and eccentricities and thus little mass is lost.

\begin{figure*}
  \centering
   \includegraphics*[scale=0.63]{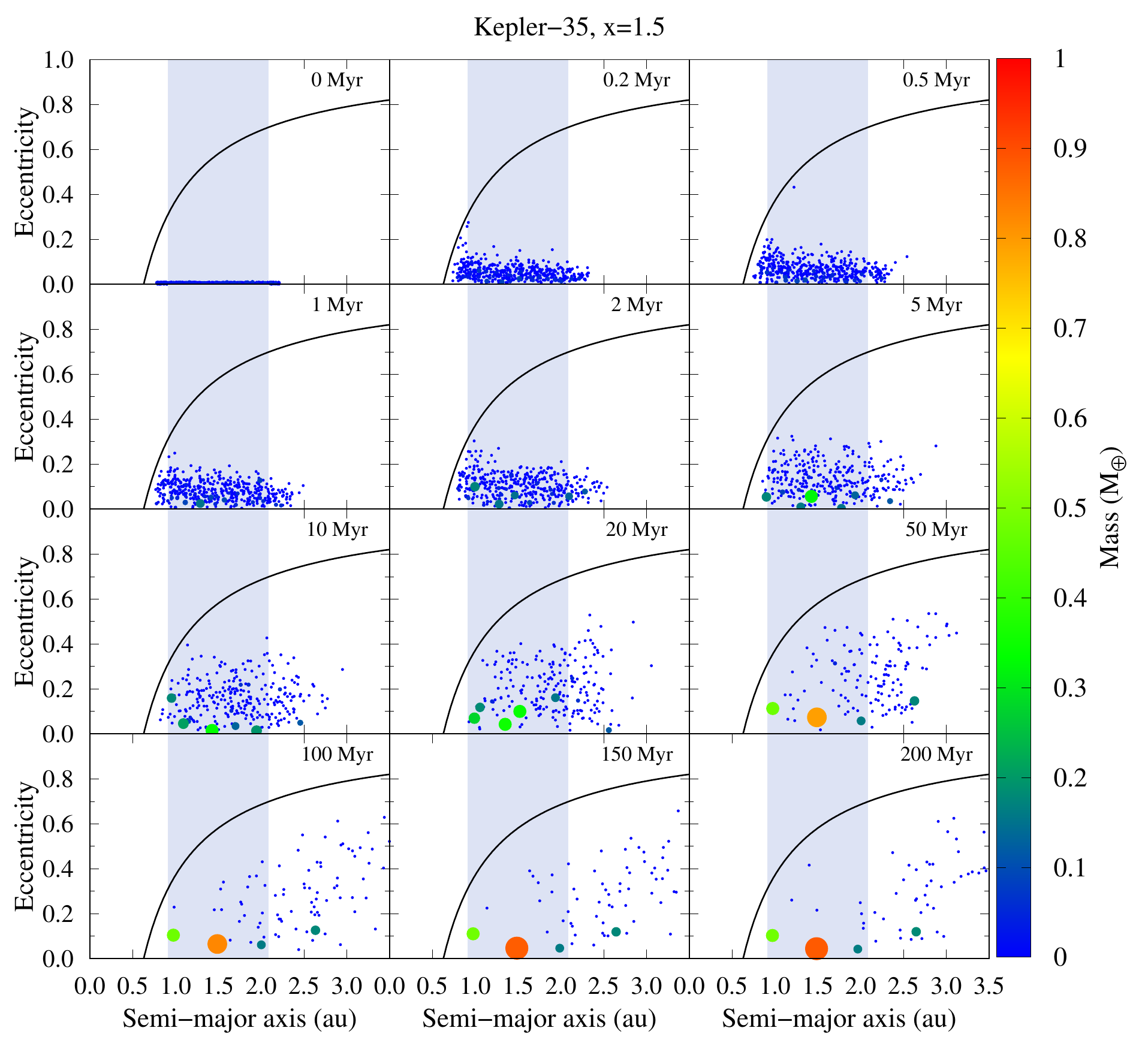}
      \caption{Snapshots in time (case \textit{Kepler-35-5}) of the dynamic evolution of embryos and planetesimals of \textit{Kepler-35} system. The horizontal and vertical axes are the semi-major axis and the eccentricity, respectively.The colored circles represent the embryos and planetesimals and their sizes are proportional to their masses and the black line is the apocenter of the host planet of the system in function of the pericenter of the particle given by $a=[a_p(1 + e_p)]/(1-e)$. The shaded region in blue represents the HZ of the system.}
         \label{k35_evolution_um}
\end{figure*}

In the case of $x=1.5$, at least one planet with mass > 0.7 $M_\oplus$ has been formed, see Figure \ref{1}. Analyzing only the bodies that were formed in the simulation inside the HZ, we have that in the \textit{Kepler-35-1} simulation, two bodies with masses of 0.6 and 0.85 $M_{\oplus}$ are formed. The \textit{Kepler-35-2} simulation is the case in which less massive bodies are formed, which are approximately 0.6 and 0.3 $M_{\oplus}$. In \textit{Kepler-35-3} the largest planet is formed, with a mass equal to 1.2 $M_{\oplus}$. The \textit{Kepler-35-4} and 5 simulations have very similar results. Where in both cases planets with about 0.85 Earth mass were formed in the simulation and with similar semi-major axis. Figure \ref{k35_evolution_um} shows snapshots in time of the dynamic evolution of the semi-major axis and the eccentricity of embryos and planetesimals of the \textit{Kepler-35-5} simulation.

\begin{figure*}
  \centering
   \includegraphics*[scale=0.63]{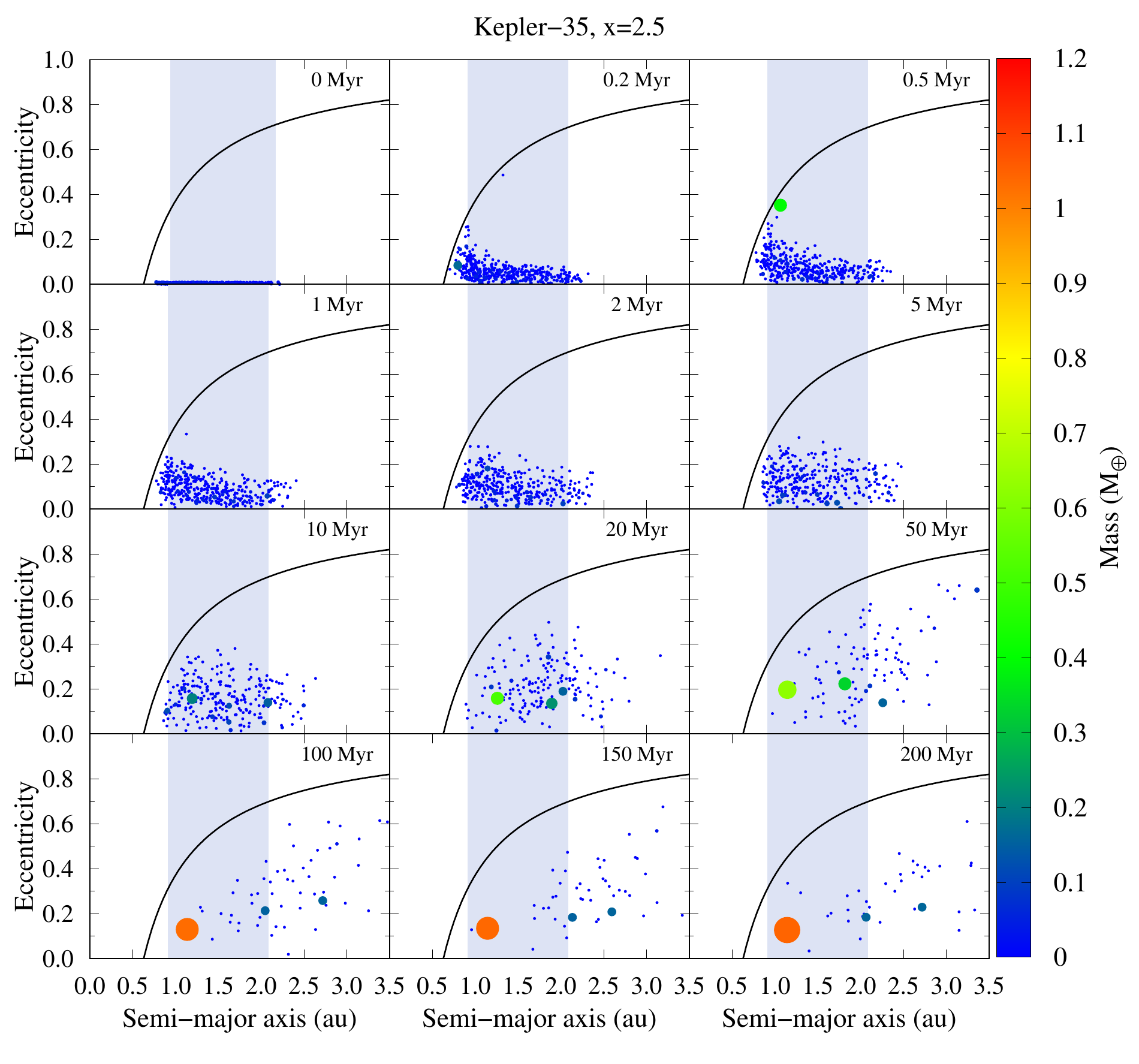}
      \caption{Snapshots in time (case \textit{Kepler-35-4}) of the dynamic evolution of embryos and planetesimals of \textit{Kepler-35} system. The horizontal and vertical axes are the semi-major axis and the eccentricity, respectively.The colored circles represent the embryos and planetesimals and their sizes are proportional to their masses and the black line is the apocenter of the host planet of the system in function of the pericenter of the particle given by $a=[a_p(1 + e_p)]/(1-e)$. The shaded region in blue represents the HZ of the system.}
         \label{k35_evolution_dois}
\end{figure*}

In the case where we have $x=2.5$, we also found that in all simulations, at least one planet with mass greater than 0.6 $M_{\oplus} $ is formed inside the HZ, see Figure \ref{2}. Looking at these planets, we have that in the \textit{Kepler-35-1} case, two planets with masses of 0.44 and 0.78 $M_{\oplus} $ are formed. In the \textit{Kepler-35-2}, the largest of them is formed, with mass equal to $1.33 M_{\oplus}$. In the \textit{Kepler-35-3} simulation, two planets with very similar masses were formed (0.65 $M_{\oplus}$). The \textit{Kepler-35-4} simulation has an Earth size with a mass of 1.04 $M_{\oplus}$, Figure \ref{k35_evolution_dois} shows snapshots in time of the dynamical evolution of the semi-major axis and eccentricity of embryos and planetesimals. Finally, in the case \textit{Kepler-35-5}, two planets are formed, the largest with a mass of 0.75 $M_{\oplus}$.

The results shown above showed that with both values of $ x $ (1.5 and 2.5) terrestrial planets were formed within the HZ. Therefore, we can conclude that this system has a great potential to house a habitable Earth type planet.

\subsubsection{Kepler-38}

The binary star system \textit{Kepler-38}, has two stars with masses of 95\% and 25\% solar masses. The brighter star is spectral class G while the secondary has spectral class M. The stars have semi-major axis 0.147 $au$ and complete an eccentric orbit around a common center of mass every 18.8 days \citep{orosz2012neptune}. The system has a planet called \textit{Kepler-38b}, known as a Neptune-sized type. Since it has not yet noticeably perturbed the stellar orbits, the planet does not have exact mass, being it an upper limit of 122 Earth masses, being its reasonable mass equal to 21 Earth masses\footnote{
This mass is found assuming that the planet follows the empirical mass-radius relationship of $M_b=(R_b/R_\oplus)^{2.06}M_\oplus$ \citep{lissauer2011architecture}. Where $M_b$ and $R_b$ are respectively the mass and the radius of the planet.}, which was used in our work as can be seen in Table \ref{parameters_systems}. The system has the most stable HZ. The planet is 0.7 $au$ away from the disk, in addition to being a planet with mass 0.016 $ M_j $, that is, the smallest of the planets studied in our work. With this, the planet causes little disturbance to the disk. At $x = 1.5$, the first ejection occurred at 7 million years and at $x=2.5$ in only 10 million years. In addition, this small disturbance causes collisions to occur in a longer timescale and consequently the formation of planets is slower. 

For $ x = 1.5 $, Figure \ref{1} shows that planets with Earth-like mass are formed in all simulations within the HZ. The \textit{Kepler-38-1} simulation is the one containing the smallest planet formed, with a mass of 0.74 $M_{\oplus}$. The \textit{Kepler-38-2, 3} and \textit{4} cases have the largest ones, with masses of 1.23, 1.14 and 1.27 $M_{\oplus}$, respectively. In addition to these three simulations having planets with similar masses, the positions in which they were formed are also similar. The \textit{Kepler-38-5} case has the planet with the largest semi-major axis, being 2.5 $au$ with mass approximately equal to 1.01 $M_{\oplus}$.

In the cases for $x = 2.5$, all simulations have at least one planet inside HZ, see Figure \ref{2}. Regarding these planets, in the \textit{Kepler-38-1} simulation, a planet with mass 1.1 $M_{\oplus}$ was formed. In the case \textit{Kepler-38-2}, one planet with mass equal to 1.26 $M_{\oplus}$ is formed in the same position as the previous case. The final configuration of the other three cases is quite similar. In these cases, three planets were formed outside the HZ and three others inside. The semi-major axis of HZ's three external are close to 1 $au$ with masses around 0.65 $M\oplus$ and the three interns have semi-major axis close to 2 $au$ with masses ranging from 0.4 to 0.65 $M_\oplus$.

From the results found, we can conclude that the system has conditions of forming and housing a habitable Earth type planet. With less massive disks on the inside (with $x=1.5$), we show that in almost all cases (except simulation 1) Earth-like planets form inside the HZ. In the cases where the simulations had more massive disks inside, only the simulations 1 and 2 Earth-type planets inside the HZ were formed.

\subsubsection{Kepler-413}

In this system, the two stars \textit{Kepler-413A} and \textit{Kepler-413B} have masses of 82\% and 54\% of the mass of the Sun. The brightest is a K dwarf spectral type while its companion a M dwarf, which results in the thinnest HZ of the systems present in our simulations.
In this system, we have the second least massive host planet of the systems treated in our work. However, unlike the Kepler-38 system that has a planet that is $\approx$ 0.6 $au$ from the disk, and therefore, resulting in a disk in which the collisions happen slowly, in this case, the planet is at the inner boundary of HZ and very close to the disk ($\approx$ 0.05 $au$), causing great disturbance. Much of the mass present in the innermost region of the disk is ejected as early as the first instants of the simulation. In the simulations where $x = 1.5$, 24\% of the total disk mass was lost in 0.5 $Myr$, while for $x = 2.5$, 38\% of the total mass was lost in the same length of time. 

In the case of $x = 1.5$, where the outermost part of the disk are conserved, thus forming more massive planets, see Figure \ref{1}. In the \textit{Kepler-413-1} simulation the largest one is formed, with mass close to 1.4 $M_{\oplus}$, but outside of HZ. In simulation \textit{2} of this system, a planet with 0.6 Earth mass is formed, also outside of HZ. In simulation \textit{3}, a planet with 1.1 $ M_\oplus$ is formed inside the HZ very close to the outer boundary. In the following two cases, \textit{Kepler-413-4} and \textit{5}, a planet was formed in each case with masses of 0.6 and 1 $M_\oplus$, respectively.

In simulations with $x=2.5$, due to the factors mentioned above, less massive planets were formed, see Figure \ref{2}. In simulation \textit{1}, one planet with 0.6 Earth mass forms far from HZ. In simulations \textit{2} and \textit{3}, two planets with similar masses, approximately 0.75 $M_\oplus$, are formed outside HZ in similar regions. In cases \textit{4} and \textit{5}, two planets also with similar masses, approximately 0.6 $M_\oplus$, are formed within the HZ.

Thus, we can conclude that the system is capable of housing habitable Earth-like planets within HZ in some cases. These cases occur when the system has a less massive embryo and planetesimal disk in its internal region ($x=1.5$).


\subsubsection{Kepler-453}

The system \textit{Kepler-453} has two stars, \textit{Kepler-453A} and \textit{Kepler-453B}. The brightest is similar to our Sun, containing 94\% of its mass, while the smaller star, Kepler-453B, has about 20\% of its mass. Stars orbit one another every 27.3 days \citep{welsh2015kepler}. This system has a giant planet that is within the HZ of the system with $\approx$ 0.8 mass of Jupiter. It is important to remember that not every planet within the HZ of a system can harbor life since the boundaries of these regions are calculated on the basis of the flow of energy received at the top of Earth's atmosphere. In calculating the stableness of the habitable region, we discovered a small band on which it was possible to distribute a disk. Thus, the beginning of the disk of material is approximately 0.85 $au$, being that the planet has semi-major axis equal to 0.790 $au$. This way the disk is very affected by the presence of the planet so close. As in previous cases, in simulations with $x = 1.5$, on average less mass was ejected. In this case 10\% of the disk mass lost to 0.5 $Myr$, and 16\% mass lost in the same time interval to $x = 2.5$. 
Although the planet is extremely close to the disk, and even within the HZ, due to its low eccentricity (0.036), not as much mass is ejected as in the case of the \textit{Kepler-34} system. Thus, planets with masses close to Earth`s mass (0.6 - 0.9 M$_\oplus$) are formed in all simulations, but all outside of HZ, see Figures \ref{1} and \ref{2}. 

\section{Final remarks}

In this work, we numerically investigate the possibility that planets with Earth mass are formed within the habitable zone of real close-binary systems with detected planets. With this, our work seeks to show which systems are capable of containing an Earth-type planet capable of harboring life.

For this, we first selected all systems with this configuration and first filtered out systems that had sufficient data for the calculation of those regions and subsequently those that had reasonable stability for a circumstellar mass disk to be distributed in those regions. Our simulations were carried for 200 $Myr$. The orbital
parameters and masses of the stellar systems were taken from the literature specified in the text.
As previously mentioned, our disks follow a bimodal configuration, composed by planetary embryos, whose mass scales along the disk with two distinct parameters, and planetesimals. In our systems. In our simulations, all bodies gravitationally interact with each other, except for the planetesimals that interact only with the other bodies, but do not interact gravitationally between them. The disks of each system extend over their habitable zone, however, not all systems have stability throughout the region, thus being partially distributed in the most stable part, as in the case of the \textit{Kepler-34, 413} and \textit{453} systems.

In order to present our results, each system has a particular evolution of the disk bodies. In the \textit{Kepler-34} system, the high eccentricity and proximity of the host planet to the HZ of the system, makes the formation of planets with mass near the Earth in the innermost parts of the disk difficult. Especially on disks where more massive embryos are in these regions because they are ejected more easily. On the other hand, disks with masses greater than 2.5 $M_\oplus$, can provide possibility of the system to harbor an Earth-like planet within the HZ. 

The \textit{Kepler-35} system, in both cases of $x$, Earth-like planets are formed within the HZ in our simulations. The host planets are not so close to the disk and their mass is also not so high, it provides enough excitation to promote collisions and consequently accretions of mass on the disk. And given the system age of approximately 8-12 $Gyr$, the existence of a planet with conditions to harbor life is possible in this system from our initial conditions. By having the broadest HZ of all systems studied in this work, and together with the high capacity to form terrestrial planets within that region, this system has the capacity to harbor one habitable planet, as can be seen in Figures \ref{1} and \ref{2}.

The \textit{Kepler-38} system, along with the \textit{Kepler-35} system, has the best ingredients for the formation of planets within the HZ. Of all the systems studied in this work, this is the one that has the least massive host planet and consequently the most stable HZ of all. For this reason, this system has the slowest planetary formation process, given by the low perturbation of the planet, resulting in a cold disk. However, we show that the formation of Earth-like planets within HZ is possible in this system in most cases for both $x$ values.

Because it has the smallest HZ of the systems, the chances of formation of planets within this region becomes smaller in the case of the \textit{Kepler-413} system. Still, planets with masses between 0.6 and 0.8 masses of Earth are formed in our simulations and in one case a planet with 0.7 $M_\oplus$ is formed within that region for $x = 2.5$. 
For $x = 1.5$, more massive planets are formed. In one of the simulations, a planet with 1 $M_\oplus$ is formed inside the HZ and another with a mass of 1.4 $M_\oplus$, but this outside the HZ.

In the \textit{Kepler-453} system, planets with mass ranging from 0.6 to 0.9 M$_\oplus$ were formed. However, none of these planets within the HZ. This occurred due to the system having a too small HZ and also by the position of the host planet within the HZ, causing the bodies to be thrown out of this region.

We can conclude from our simulations, that terrestrial planets can be formed in some known CB systems. To date, no such planet has been detected in this type of system. In addition, it is shown that these planets are within the habitable zone of the systems. Knowing the location where these planets may be located, and mainly that their existence is possible, from an observational point of view, more precise searches can be performed on the systems shown. Another important result that we show is that some systems at the end of the stability test, had coorbital bodies to their host planets. This can be seen in Figure \ref{stability_final} in the systems \textit{Kepler-16, 47, 453} and \textit{1647}. With that, if the planet is within the HZ, which occurs in these systems mentioned, the coorbital bodies will also be. In other words, this result shows that even if a gaseous giant planet, unable to harbor life as we know it, is within HZ, if an Earth-type planet is sharing the same orbit as it, it can still be habitable.

\section*{Acknowledgements}
The authors thank an anonymous reviewer whose comments greatly improved the manuscript. We also thank Paul Mason who helped us with his code for calculating habitable zones and Rafael Sfair with all computational support.

The work was carried out with the support of the \textit{Improvement Coordination
Higher Education Personnel} - Brazil (CAPES) - Financing Code 001 and National Council for Scientific and Technological Development (CNPq, proc. 305210/2018-1 and 310317/2016-9). A. I. acknowledges support from FAPESP via grants 16/19556-7 and 16/12686-2, and CNPq via grant 313998/2018-3.
The research had computational resources provided by the thematic projects 
FAPESP proc. 2016/24561-0, FAPESP proc. 2015/50122-0 and the \textit{Center for Mathematical Sciences Applied to Industry (CeMEAI)}, funded by FAPESP (proc. 2013/07375-0).
\section*{ORCID iDs}
G. O. Barbosa \orcidicon{0000-0002-1147-2519} \href{https://orcid.org/0000-0002-1147-2519}{https://orcid.org/0000-0002-1147-2519}\\
O. C. Winter \orcidicon{0000-0002-4901-3289} \href{https://orcid.org/0000-0002-4901-3289}{https://orcid.org/0000-0002-4901-3289}\\
A. Amarante \orcidicon{0000-0002-9448-141X} \href{https://orcid.org/0000-0002-9448-141X}{https://orcid.org/0000-0002-9448-141X}\\
A. Izidoro \orcidicon{0000-0003-1878-0634} \href{https://orcid.org/0000-0003-1878-0634}{https://orcid.org/0000-0003-1878-0634}\\
R. C. Domingos \orcidicon{0000-0002-0516-0420} \href{https://orcid.org/0000-0002-0516-0420}{https://orcid.org/0000-0002-0516-0420}\\
E. E. N. Macau \orcidicon{0000-0002-6337-8081} \href{https://orcid.org/0000-0002-6337-8081}{https://orcid.org/0000-0002-6337-8081}
\bibliographystyle{mnras}
\bibliography{binarios} 
\bsp	
\label{lastpage}
\end{document}